\documentclass[prl,twocolumn,floatfix,superscriptaddress]{revtex4-1}
\usepackage{dcolumn,amsmath}
\usepackage[dvips]{graphicx}
\usepackage{bm}
\usepackage{hyperref}

\newcommand{\eEDM}{{\em e}EDM}

\begin{document}
 \title{Verification of g-factors for lead monofluoride ground state, PbF}
\author{L.V.\ Skripnikov}\email{leonidos239@gmail.com}
\author{A.N.\ Petrov}\email{alexsandernp@gmail.com}
\author{A.V.\ Titov}
\homepage{http://www.qchem.pnpi.spb.ru}
\affiliation{National Research Centre ``Kurchatov Institute'' B.P. Konstantinov Petersburg Nuclear Physics Institute, Gatchina, Leningrad district 188300, Russia}
\affiliation{Dept.\ of Theoretical Physics, St.Petersburg State University, 198504, Russia}
\author{R.J.\ Mawhorter}
\author{A.L.\ Baum}
\affiliation
{Department of Physics and Astronomy, Pomona College, Claremont, California 91711-6359, USA}
\author{T.J.\ Sears}
\affiliation{Chemistry Department, Brookhaven National Laboratory, Upton, New York 11973-5000, USA}
\author{J.-U.\ Grabow}
\affiliation{Gottfried-Wilhelm-Leibniz-Universität, Institut für Physikalische Chemie and Elektrochemie, Lehrgebiet A, D-30167 Hannover, Germany}
\begin{abstract}
We report the results of our theoretical study and analysis of earlier experimental data for the g-factor tensor components of the ground $^2\Pi_{1/2}$ state of free PbF radical. The values obtained both within the relativistic coupled-cluster method combined with the generalized relativistic effective core potential approach and with our fit of the experimental data from [R.J. Mawhorter, B.S. Murphy, A.L. Baum, T.J. Sears, T. Yang, P.M. Rupasinghe, C.P. McRaven, N.E. Shafer-Ray, L.D. Alphei, J.-U. Grabow, Phys. Rev. A 84, 022508 (2011); A. Baum, B.S. thesis, Pomona College, 2011].
The obtained results agree very well with each other but contradict the previous fit performed in the cited works. Our final prediction for g-factors is $G_{\parallel}= 0.081(5)$, $G_{\perp}=-0.27(1)$.
\end{abstract}

\maketitle


\section{Introduction}

Lead monofluodide, PbF, molecule is one of prospective systems to search for the electron electric dipole moment (\eEDM). It was studied and discussed during three decades in many papers including \cite{Kozlov:87, Dmitriev:92, Shafer-Ray:06, Shafer-Ray:08E, Baklanov:10, Petrov:13}.
It was recently shown in Ref.~\cite{Alphei:11} that some ``enhanced'' (coincidental) near-degeneracy for the levels of opposite parity in the ground rotational state $J=1/2$ for $^{207}$PbF of the ground electronic state $^2\Pi_{1/2}$ \cite{Shafer-Ray:08E}  takes place that is caused by the near cancellation between the shifts in the energies of these levels due to omega-type doubling and the magnetic hyperfine interaction. This can lead to suppression of systematic errors in an experiment.

In Ref. \cite{Skripnikov:14c} we have calculated the parameters (more generally, the characteristics of atoms in compounds~\cite{Lomachuk:13,Titov:14a,Skripnikov:15b}]) required to interpret the experimental energy shift in terms of the \eEDM\ and other effects of simultaneous violation of space parity (P) and/or time-reversal invariance (T) including the P-odd anapole moment \cite{Alphei:11} and the T,P-odd pseudoscalar-scalar electron-nucleus neutral current
interaction for the ground $^2\Pi_{1/2}$ state.  For instance, the effective electric field in PbF was found to be greater than or equal to those in the other transition element compounds considered (1.7 times larger than in  HfF$^+$~\cite{Petrov:07a, Fleig:13}, 1.4 larger than in PtH$^+$~\cite{Skripnikov:09}, and 1.1 larger than in WC~\cite{Lee:13a} and TaN~\cite{Skripnikov:15c})).

In the present paper our aim is to study the PbF g-factor for the $^2\Pi_{1/2}$ term which is required for preparation of experiments on the molecule \cite{Shafer-Ray:06,Yang:13,Petrov:14}. Up to now the g-factors have been measured in Ref.~\cite{Mawhorter:11,Baum:11} only. Previous theoretical estimations and calculations of g-factors have been performed in Refs.~\cite{Kozlov:87, Dmitriev:92, Baklanov:10}.

\section{Molecular Hamiltonian}

We represent the molecular Hamiltonian for $^{208}$PbF as~\cite{Petrov:13}:
\begin{equation}
{\rm \bf H}_{\rm mol} = {\rm \bf H}_{\rm rot} + {\rm \bf H}_{\rm hfs} + {\rm \bf H}_{1} + {\rm \bf H}_{ext} .
\end{equation} 
Here ${\rm \bf H}_{\rm rot}$ is the rotational Hamiltonian and ${\rm \bf H}_{\rm hfs}$ is the hyperfine interaction between electrons and nuclei. ${\rm \bf H}_{1}$ includes the nuclear spin -- rotational interaction and also effectively takes into account the rotational and hyperfine interactions between $^2\Pi_{1/2}$ and other electronic states. 
${\rm \bf H}_{ext}$ describes the interaction of the molecule with an external magnetic field $\mathbf{B}$.
Parameters for ${\rm \bf H}_{\rm rot}$, ${\rm \bf H}_{\rm hfs}$, and ${\rm \bf H}_{1}$ are taken from Ref. \cite{Petrov:13}.
For ${\rm \bf H}_{ext}$ we have:
\begin{equation}
 \label{Bext}
{\rm \bf H}_{ext} = \mu_{B}~\mathbf{B}\cdot\widehat{G}\cdot \mathbf{S}^{\prime} -g_1\mu_{N}~\mathbf{B}\cdot \mathbf{I_1}
\end{equation}
Here $\mathbf{S}^{\prime}$ is effective spin defined by the following equations: $\mathbf{S}^{\prime}_{\hat{n}}|\Omega> = \Omega|\Omega>$, $\mathbf{S}^{\prime}_{\pm}|\Omega=\mp 1/2> = |\Omega=\pm 1/2>$,
$\mathbf{S}^{\prime}_{\pm}|\Omega=\pm 1/2> = 0$ \cite{Kozlov:87, Dmitriev:92}, $\mathbf{I_1}$ is the
angular-momentum operator of the fluorine nuclei,
$\mu_{B}$ and $\mu_{N}$ are Bohr and nuclear magnetons respectively, and $g_1=5.25773$ is the $^{19}$F
nuclear $g-$factor.

In the molecular frame coordinate system the tensor contractions
\begin{gather}
\label{contraction}
\mathbf{B} \cdot\widehat{G}\cdot\mathbf{S}^{\prime}=G_{||}\mathbf{B}_{0}\mathbf{S}_{0}^{\prime}-G_{\perp}(\mathbf{B}_{1}\mathbf{S}_{-1}^{\prime}
+\mathbf{B}_{-1}\mathbf{S}_{1}^{\prime}) 
\end{gather}
are determined by the body-fixed $g-$factors
\begin{eqnarray}
 \label{Gpar}
  G_{\parallel} &=&\frac{1}{\Omega} \langle \Psi_{^2\Pi_{1/2}} |\hat{L}^e_{\hat{n}} - g_{S} \hat{S}^e_{\hat{n}} |\Psi_{^2\Pi_{1/2}} \rangle,  
\end{eqnarray}
\begin{eqnarray}
 \label{Gperp}
G_{\perp} &=&\langle \Psi_{^2\Pi_{1/2}} |\hat{L}^e_{+} - g_{S} \hat{S}^e_{+} |\Psi_{^2\Pi_{-1/2}} \rangle,  
\end{eqnarray}
where ${\vec{L}}^e$ and ${\vec{S}}^e$ are the electronic orbital and electronic spin momentum operators, respectively; $g_{S} = -2.0023$ is a free$-$electron $g$-factor; $\hat{n}$ is the unit vector along the molecular axis directed from Pb to F.

In this paper the parameters  $G_{\parallel}$ and $G_{\perp}$ are obtained
(i) by using 
Eqs.~(\ref{Gpar},\ref{Gperp}) from calculation of the {\it electronic} wavefunction  $\Psi_{^2\Pi_{1/2}}$
and
(ii) by fitting the experimentally observed transitions reported in Ref.~\cite{Baum:11}. 

\section{Methods}

The matrix elements (\ref{Gpar},\ref{Gperp}) were calculated using the computational scheme similar to that used by us in Ref.~\cite{Skripnikov:14c}. The basis set for Pb was taken from Ref.  \cite{Skripnikov:14c}. For F the aug-ccpVQZ basis set \cite{Kendall:92} with two removed g-type basis functions was employed. The Pb$-$F internuclear distance was set to 3.9 a.u., which is close to the experimental datum 
3.8881(4) a.u. \cite{Lumley:77}, which was later confirmed by Ref.~\cite{Ziebarth:98}.
Inner core $1s-4f$ electrons of lead were excluded from the correlation calculation using the  ``valence'' semi-local version of the generalized relativistic effective core potential (GRECP) approach \cite{Mosyagin:10a,Titov:99}. Note that the approach allows one to account for the Breit interaction very effectively \cite{Petrov:04b, Mosyagin:06amin, Mosyagin:10a}. All the other 31 electrons were included into the calculation.
Electron correlation effects were considered within the relativistic two-component coupled-cluster approach with accounting for single and double cluster amplitudes, CCSD, as well as single, double and perturbative triple cluster amplitudes, CCSD(T).
Note that the matrix element (\ref{Gperp}) is off-diagonal.
Therefore, it was calculated within the linear-response two-component coupled-cluster method with single and double cluster amplitudes \cite{Kallay:5}. 
The coupled-cluster calculations were performed using the {\sc dirac12} \cite{DIRAC12} and {\sc mrcc} \cite{MRCC2013} codes. 
Matrix elements of the operators corresponding to (\ref{Gpar},\ref{Gperp}) over the molecular spinors were calculated with the code developed in Refs. 
\cite{Skripnikov:11a,Skripnikov:13b,Skripnikov:13c,Skripnikov:14b,Skripnikov:15b,Petrov:14}.

To obtain the experimental values for $G_{\parallel}$ and $G_{\perp}$ we have performed two fits using the data from Ref. \cite{Baum:11}. In ``fit~1'' the Zeeman shifts of $J = 1/2$ to $J = 3/2$ transitions for the ground vibrational level of $^2\Pi_{1/2}$ electronic state are obtained by numerical diagonalization of the molecular Hamiltonian (${\rm \bf H}_{\rm mol}$) on the basis set of the electronic-rotational wavefunctions. The scheme of the calculation is similar to that employed in Refs. \cite{Petrov:11, Petrov:13,Lee:13a}. Only the $G_{\parallel}$ and $G_{\perp}$ parameters were optimized. The other parameters of ${\rm \bf H}_{\rm mol}$ were taken from Ref.~\cite{ Petrov:13}.
In ``fit~2'' we have reproduced the scheme described in Ref.~\cite{Mawhorter:11}.

\section{Results and discussion}

The results of our calculations  of g-factors for the PbF ground state together with the results of previous studies are given in Table \ref{TResults}. One can see that the value of $G_{\parallel}$ is stable with respect to improvement of the electron correlation treatment in the present study (from CCSD to CCSD(T) level).

\begin{table}[!h]
\caption{Calculated values of g-factors ($G_{\parallel}$,   $G_{\perp}$) of the $^2\Pi$ state of PbF.
}
\label{TResults}
\begin{tabular}{ l  c  c}
\hline\hline
 Method                             &   $G_{\parallel}$ & $G_{\perp}$ \\
\hline      
SCF $^a$, \cite{Kozlov:87}       & 0.034 $< G_{\parallel} <$ 0.114 & -0.438   $< G_{\perp} <$  -0.269     \\ 
SCF $^a$, \cite{Dmitriev:92}     &   0.114 & -0.438     \\ 
13e-SODCI$^b$, \cite{Baklanov:10}&   0.082 & -0.319     \\ 
\\

 31e-CCSD, this work                            &   0.081 & -0.274     \\ 
  31e-CCSD(T),                         &   0.081 & ---        \\
this work  &&         \\    
\\
Experiment, \cite{Mawhorter:11}  &   0.12  &  -0.38     \\  
  Experiment + fit~1,            &   0.081 &  -0.269  \\  
this work  &&         \\  
  Experiment + fit~2,            &   0.085 &  -0.271  \\  
this work  &&         \\

\hline\hline
\end{tabular}
\\
$^a$ SCF, self consistent field.
\\
$^b$ 13-electron SODCI, spin-orbit direct configuration interaction, \cite{Baklanov:10}.
Outer-core electrons $5s^25p^65d^{10}$ of Pb are excluded from the correlation treatment.
\\
\end{table}

$G_{\parallel}$ and $G_{\perp}$ values obtained by fit~1 and fit~2 (see Methods section) are also given in Table~\ref{TResults}. The deviations of our fits from the observed Zeeman shifts are given in Table~\ref{spectr208}. For the last seven transitions the shifts are reproduced with deviations which are much larger than the declared experimental accuracy. One is inclined to suspect that the accuracy is overestimated for these transitions. We note however, that the experimental $(\Delta U/B)_{\rm obs}$
 values for Zeeman components that only differ (model independent) in sign (e.g. F$_L$, MF$_L$ $\to$ F$_U$, MF$_U$ = 1, 0 $\to$ 2, 1 vs. 1, 0 $\to$ 2, -1; 1, 1 $\to$ 2, 2 vs. 1, -1 $\to$ 2, -2; 1, 1 $\to$ 2, 0 vs. 1, -1 $\to$ 2, 0; 0, 0 $\to$ 1, 1 vs. 0, 0 $\to$ 1, -1) agree within their error bars, which indicates correct accuracy estimations.  It is also the case that the deviations for those pairs are systematic and not statistical.  It seems that the F$_L$ $\to$ F$_U$ = 1 $\to$ 2 pattern is predicted to be somewhat too narrow while the F$_L$ $\to$ F$_U$ = 0 $\to$ 1 pattern is somewhat too wide.

We also note that the $G_{\parallel}$ = 0.085, $G_{\perp}$ = -0.271 parameters obtained in fit 2 differ substantially from the $G_{\parallel}$ = 0.12, $G_{\perp}$ = -0.38 values obtained by the same method and reported in the Ref.~\cite{Mawhorter:11}. Our results here show good agreement between $G_{\parallel}$ and $G_{\perp}$ obtained in fit 1, fit 2, and the {\it ab~initio} calculation.  While both Ref.~\cite{Mawhorter:11} values are higher by a common factor of $\sim$1.4--1.5,  the origin of the discrepancies is not clear at present and will require further investigation.
 
Our final values for the g-factors are $G_{\parallel}$ = 0.081(5) and $G_{\perp}$ = -0.27(1). It should be noted that these smaller g-factor values and their improved accuracy together favor the
experimental search for the electron electric dipole moment and other parity-violating and related effects \cite{Borschevsky:13,Flambaum:2013} in PbF due to the additional suppression of systematic errors.

\begin{table}
\caption{Observed Zeeman shifts $(\Delta U/B)_{\rm obs}$ (MHz/Gauss) of the $J = 1/2$ to $J = 3/2$ transitions for $^{208}$Pb$^{19}$F \cite{Baum:11}. The number in parenthesis gives two standard deviation error of the final digits of precision. The subscripts $U$ and $L$  refer to the upper and lower energy level of the transition, respectively. F is the total angular momentum of PbF, MF is its projection on laboratory axis. The deviation of $n$-th fit  is given by $\delta_n= (\Delta U/B)_{\rm fit} - (\Delta U/B)_{\rm obs}$
in units of the last digit of precision.}
\begin{tabular}{cccccccc}
  Unsplit line (MHz)  & F$_L$ & F$_U$ & MF$_L$ & MF$_U$ & $(\Delta U/B)_{\rm obs}$ \cite{Baum:11} & $\delta_1$  &$\delta_2$   \\
\hline
 18414.588 & 1 & 2 & -1 & -1 &  0.0665(13)   & -40  &  -30   \\
           &   &   &  0 &  0 & -0.00050(93)  & 107  &   50   \\
           &   &   &  1 &  1 & -0.0635(13)   &  9   &  0     \\
 18462.193 & 0 & 1 &  0 &  0 &  0.00032(90)  & -89  & -32    \\
           &   &   &  0 & -1 & -0.1369(30)   & -4   &  5     \\
           &   &   &  0 &  1 &  0.1363(29)   & -2   &  11    \\
 18497.136 & 1 & 1 & -1 & -1 &  0.00766(86)  & -150 & -70    \\
           &   &   &  1 &  1 & -0.00729(92)  &  119 &  33    \\
           &   &   &  1 &  0 & -0.1428(17)   &  0   & -15    \\
           &   &   &  0 & -1 & -0.1328(21)   & -33  & -46    \\
           &   &   &  0 &  1 &  0.1345(13)   &  29  &  29    \\
           &   &   & -1 &  0 &  0.1427(9)    &   2  &  16    \\
 22574.934 & 1 & 2 & -1 & -1 & -0.03864(27)  &  28  &  100   \\
           &   &   &  0 &  0 & -0.00005(90)  &  2   &  5     \\
           &   &   &  1 &  1 &  0.03851(9)   & -14  & -87    \\
           &   &   &  1 &  0 &  0.1023(36)   &  64  &  62    \\
           &   &   &  0 & -1 &  0.07296(25)  & -264 & -207   \\
           &   &   & -1 & -2 &  0.03411(60)  & -211 & -85    \\
           &   &   &  1 &  2 & -0.03406(49)  &  206 &  80    \\
           &   &   &  0 &  1 & -0.07267(52)  &  229 &  178   \\
           &   &   & -1 &  0 & -0.10323(79)  & -546 & -529   \\
 22691.931 & 0 & 1 &  0 & -1 &  0.11114(48)  &  371 &  433   \\
           &   &   &  0 &  1 & -0.11133(41)  & -346 & -414   \\
\hline
\end{tabular}
\label{spectr208}
\end{table}


\section*{Acknowledgement}

The molecular calculations were partly performed at the Supercomputer ``Lomonosov''.
This work is supported by the SPbU Fundamental Science Research grant from Federal budget No.~0.38.652.2013 and RFBR Grant No.~13-02-01406. L.S.\ is also grateful to the grant of President of Russian Federation No.MK-5877.2014.2 and Dmitry Zimin ``Dynasty'' Foundation. 
J.-U.G. acknowledges funding from the Deutsche Forschungsgemeinschaft
(DFG) and the Land Niedersachsen, and R.J.M.  and A.L.B. are grateful for research support provided by the Pomona College Sontag Fellowship Program.


\end{document}